\documentstyle[twocolumn,aps,psfig]{revtex}
\begin{document}
\def\r{{\bf{r}}}
\def\k{{\bf{k}}}
\def\K{{\bf{K}}}
\def\q{{\bf{q}}}
\def\Q{{\bf{Q}}}
\def\p{{\bf{p}}}
\def\P{{\bf{P}}}

\title{Nonlocal Dynamical Correlations of Strongly Interacting Electron Systems}
\author{M. H. Hettler$^1$, A. N. Tahvildar-Zadeh$^1$, M. Jarrell$^1$, T. Pruschke$^2$, and H. R. Krishnamurthy $^3$}
\address{$^1$ Department of Physics, University of Cincinnati, Cincinnati,
OH 45221}
\address{$^2$ Institut f\"ur Theoretische Physik, Universit\"at Regensburg,
Regensburg, Germany}
\address{$^3$Department of Physics, IISc,  Bangalore 560012, India}
\date{\today}
\maketitle
\begin{abstract}
We introduce an extension of the dynamical mean field approximation (DMFA)
which retains the causal properties and generality of the DMFA, but allows 
for systematic inclusion of non-local corrections.  Our technique maps the 
problem to a self-consistently embedded cluster.  The DMFA (exact result) 
is recovered as the cluster size goes to one (infinity).   As a demonstration, 
we study the Falicov-Kimball model using a variety of cluster sizes.  We 
show that the sum rules are preserved, the spectra are positive definite, 
and the non-local correlations suppress the CDW transition temperature.
\end{abstract}

\narrowtext
{\it Introduction.} Strongly interacting electron systems have been on the 
forefront of theoretical and experimental interest for several decades.
This interest has intensified with the discovery of a variety
of Heavy Fermion and related non Fermi liquid systems and the high-$T_c$ 
superconductors. In all these systems strong electronic interactions play 
a dominant role in the selection of at least the low temperature phase.
The simplest theoretical models of strongly correlated electrons, the 
Hubbard model (HM) and the periodic Anderson model (PAM), have remained 
unsolved in more than one dimension despite a multitude of sophisticated 
techniques introduced since the inception of the models.\par \indent
With the ground breaking work by Metzner and Vollhardt \cite{metzvoll}
it was realized that these models become  significantly  simpler
in the limit of infinite dimensions, $D=\infty$. Namely, provided that 
the kinetic energy is properly rescaled as $1/\sqrt{D}$, they retain only 
local, though nontrivial dynamics: The self energy is constant in momentum 
space, though it has a complicated frequency dependence. Consequently,
the HM and PAM  map  onto a generalized single impurity Anderson model.
The thermodynamics and phase diagram have been obtained numerically
by quantum Monte Carlo (QMC) and other methods. \cite{jarrell,jarrell2,lisa}\par \indent
The name dynamical mean field approximation (DMFA) has been coined for 
approximations in which a purely local self energy (and vertex function) 
is assumed in the context of a finite dimensional electron system. 
While it has been shown that this 
approximation captures many key features of strongly correlated systems
even in a finite dimensional context, the DMFA, which leads to an 
effective single site theory, has some obvious limitations.  For example, 
the DMFA can not describe phases with explicitly nonlocal order parameters,
such as d--wave superconductivity, nor can it describe the short-ranged
spin correlations seen in the metallic state.  Consequently, there have 
been efforts to extend DMFA by inclusion of nonlocal correlations, which 
would correspond to $1/D$--corrections to the self energy 
of the $D=\infty$ models\cite{peter1,avi}.  These attempts have been
only partially successful because of the difficulties of formulating a 
causal\cite{note1} theory out of nonlocal Green functions. The nonlocal 
Green functions do not have a negative-definite imaginary part, so any self 
energy diagram constructed with them is not guaranteed to preserve 
causality.  In fact, in the work by Schiller and Ingersent\cite{avi} 
on the Falicov--Kimball model (FKM) violations of the spectral sum rule
occurred for moderately large values of the interaction strength.\par \indent
In this work we introduce a new method that includes short ranged dynamical
correlations and allows for nonlocal order parameters.  The method is an 
iterative self-consistency scheme on a finite size cluster with periodic 
boundary conditions. The essential approximation is the assumption that the 
self energy is only weakly momentum dependent so that it is well approximated
on a coarse grid of cluster momentum points $\K$. This approximation will be 
very 
good in high dimensions, but in low dimensions its validity 
is less clear.  However, in many correlated systems, 
the momentum dependence of the
self energy is believed to be less important than its energy dependence,
since the physical properties are dominated by a weakly dispersive
feature in the electronic spectra near the Fermi surface, as seen, e.g.,
in experiments on Heavy--Fermion systems.\cite{exper} \par \indent
The paper is organized as follows: First, we briefly review 
the DMFA, which is reproduced by our method if we choose
a cluster consisting of only a single site. We then describe 
the new technique which we name dynamical cluster approximation (DCA).
Finally, we demonstrate the method by example of the Falicov-Kimball model.\\

{\it Dynamical Mean Field Approximation. } The DMFA assumes that the
self energy is a purely local functional of the local Green function only,
$\Sigma_{i,j} = \Sigma_{i,i}(G_{i,i}) \delta_{i,j}$. Consequently, the self 
energy has no momentum dependence, and the lattice problem may be mapped
onto a self-consistently embedded impurity problem.  The resulting DMFA
algorithm has the following steps:
(1)  The procedure starts with a local host Green function $\cal{G}$ that 
includes self energy processes at all lattice sites except at 
the ``impurity'' site {\bf i} 
under consideration.  $\cal{G}$ defines the undressed Green function of
a generalized Anderson impurity model which is then solved by some 
technique, e.g. the  QMC-method.  
(2)  Then $\Sigma_{i,i}= {\cal{G}}^{-1} - G^{-1}_{imp}$, where $G_{imp}$ is 
the computed Green function of the generalized Anderson impurity model. 
(3) This self energy is assumed to be also the self energy of the lattice. 
Consequently, the local lattice Green function follows from 
$G_{i,i}={1\over N} \sum_{{\k}} (G_{o}^{-1}(\k) - \Sigma_{i,i})^{-1}$, 
where $G_{o}(\k)$ is the bare lattice Green function and $N$ is the 
(infinite) number of points of the lattice.
(4)  The iteration loop closes by defining the new
${\cal{G}}^{-1} =G_{i,i}^{-1} +\Sigma_{i,i}$. 
The iteration typically continues until $G_{i,i}=G_{imp}$ to within the desired
accuracy, and the procedure may be shown to be completely causal.\par \indent

{\it Dynamical Cluster Approximation} (DCA).  We consider a cluster
of size $N_c=L^D$ with periodic boundary conditions. 
The corresponding first Brillouin 
zone is divided into $N_c$ cells of size $(2\pi/L)^D$. 
The algorithm begins with a guess, usually zero, for the 
cluster self energy $\Sigma_c(\K)$ (here and in the following 
we suppress the frequency argument).
We now define a Green function $\bar{G}$ as
\begin{equation}
\bar{G}(\K) =\frac{N_c}{N} \sum_{\k'} (z-\epsilon_{\K+\k'} +\mu- 
\Sigma_c(\K))^{-1}
\,,
\label{gbar}
\end{equation}
where the $\k'$ summation runs over the momenta of the cell about the 
cluster momentum  $\K$. $z$ is the (complex) frequency argument, $\mu$ the
chemical potential.$\bar{G}$ is causal provided that its 
proper self energy $\Sigma_c(\K)$ is causal.
It is a coarse grained average of the lattice Green function in momentum 
space with a self energy  $\Sigma_c(\K)$.  Before a new estimate for the
self energy can be formulated, we calculate the host cluster propagator 
$\cal{G}(\K)$ using
\begin{equation}
{\cal{G}}^{-1}(\K)= \bar{G}^{-1}(\K)+ \Sigma_c(\K).
\label{gsck}
\end{equation}
This is the ``cluster exclusion'' to prevent over-counting of 
self energy diagrams on the cluster.  Since the self energy in 
Eq.~\ref{gbar} is independent of the integration variable, Eqs.~\ref{gbar} 
and \ref{gsck} are formally identical to the corresponding equations 
(steps 3 and 4) used in the DMFA (after rescaling $\k'$).  Thus, at this 
point the DCA is equivalent to $N_c$ independent DMFA's, one for 
each  $\K$.
That Eq.\ \ref{gsck} preserves causality can be seen as follows:
Since $\Sigma_c(\K)$ in Eq.\ \ref{gbar} does not depend
on $\k'$, the sum on $\k'$ can be rewritten as an energy integral with a 
$\K$-dependent density of states (DOS) $\rho_{\K}(\epsilon)$. However, for any positive 
semi--definite, normalized function $\rho_{\K}(\epsilon)$ one has
$
\int d\epsilon\frac{\rho_{\K}(\epsilon)}{z+\mu-\Sigma_c(\K)-\epsilon}=
(z-\Sigma_c(\K)-\varepsilon_{\K}+\mu-\Gamma_{\K}(z+\mu-\Sigma_c(\K)))^{-1}
$
with an effective ``dispersion'' 
$\varepsilon_{\K}
=\frac{N_c}{N}\sum_{\k'}\epsilon_{\K+\k'}
$
for the embedded cluster and a causal function 
$\Gamma_{\K}(z+\mu-\Sigma_c(\K))$ \cite{comment}, which is the 
self energy of ${\cal{G}}$.

\begin{figure}[h]
\leavevmode\centering\psfig{file=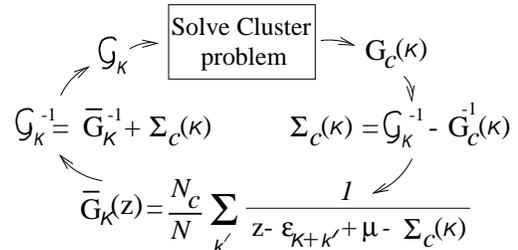,width=2.6 in}
\caption{ Schematic sketch of the dynamical cluster algorithm.
}
\label{algorithm}
\end{figure}

Given a causal host cluster propagator $\cal{G}(\K)$ we then compute the 
interacting cluster Green function $G_c(\K)$ (or self energy $\Sigma_c$) 
by any convenient method.  This introduces non-local interactions and 
correlations between the different momentum cells.  
$\Sigma_c(\K)$ is obtained via
\begin{equation}
\Sigma_c(\K)={\cal{G}}^{-1}(\K) - G^{-1}_c(\K).
\label{sigc}
\end{equation}
$\Sigma_c(\K)$ is assumed to be a good approximation of the lattice self 
energy at the cluster momenta.  It is fed into Eq. \ref{gbar} to generate 
a new $\bar{G}(\K)$.  This process is repeated until upon convergence of 
the algorithm $\bar{G}(\K)= G_c(\K)$. The schematics of the algorithm is
sketched in Fig. \ref{algorithm}.

{\it Discussion of the algorithm.} 
Several assumptions were made in the construction of this algorithm.
The first is the weak momentum dependence of the self energy which is 
equivalent to assuming that the dynamical intersite correlations have 
some short spatial range $b \alt L/2$.  Then, according to Nyquist's 
sampling theorem\cite{nyquist}, 
to reproduce these correlations in the self energy, we need only sample 
the reciprocal space at an interval of $\Delta k\approx 2\pi/L$;
i.e., on a cluster of $N_c=L^D$ points within the first Brillouin zone.
Equivalently, $\Sigma(\K+\k')\approx \Sigma(\K)$ for each $\k'$
within a cell of size $\left( \pi/b\right)^D$, so
the lattice self energy is well approximated by the self 
energy $\Sigma_c(\K)$ obtained from the coarse-grained cluster. 
Thus, the algorithm is a natural extension of the DMFA.
The second assumption is the form of Eq.~\ref{gbar}. 
This choice is not unique, but it is the simplest that maintains
causality and produces an algorithm that both recovers the DMFA 
when $N_c=1$ and becomes exact when $N_c=\infty$. When $N_c=1$,
the $\k'$ summation runs over the complete Brillouin zone and 
$\bar{G}$ is the local Green function.  When $N_c=\infty$, the $\k'$ 
summation vanishes.
\par \indent
We want to stress that the DCA is a general scheme not specialized
to a  particular model 
of interest or to the technique used to obtain 
the cluster self energy.  A variety of techniques, including perturbation 
theory\cite{dcapt} 
(NCA, the fluctuation exchange approximation\cite{bickers}), 
quantum Monte Carlo, or numerical renormalization group can also be used 
to solve the embedded cluster problem.\\

{\it Application: The Falicov--Kimball model.}
The spinless FKM can be considered as a simplified Hubbard model in which 
one spin species is prohibited to hop and has consequently only local 
dynamics. The Hamiltonian reads
\begin{equation}
H = - t \sum_{<i,j>} d^{\dagger}_i d_j -\mu \sum_i 
(n^d_i +n^f_i) + 
U  \sum_i n^d_i n^f_i
\end{equation} 
with $n^d_i = d^{\dagger}_i d_i$, $n^f_i = f^{\dagger}_i f_i$, and 
in the particle--hole symmetric case  which we consider, $\mu=U/2$.
We measure 
energies in units of the hopping element $t$.  For $D \ge 2$ the system has a 
phase transition from a homogeneous high temperature phase with  
$\langle n^d_i\rangle =\langle n^f_i\rangle =1/2$ to a checkerboard phase 
(a charge density wave with ordering vector ${\bf{Q}}=(\pi,\pi,...)$) with  
$\langle  n^d_i\rangle \ne \langle n^f_i\rangle$ for $0 < U < \infty$. 
\cite{brandt1}
In contrast to the Hubbard and related models, within the DCA the FKM
can be solved without the application of QMC 
because the f-electrons are static, acting as 
an annealed disorder potential on the dynamic d-electrons.
We generalize the algorithm of Brandt and Mielsch \cite{brami}
to a finite size cluster. Given an initial host
Green function ${\cal{G}}_{ij}$ of the d-electrons, the algorithm first 
computes the Boltzmann weights $w_f$ of all configurations $\{f\}$ 
of f-electrons on the cluster, as $w_f=w_f^0/Z$ with
\begin{eqnarray}
w_f^0= 2^{Nc}\prod_{\omega_n} \det \frac{{\cal{G}}^{-1}_{ij}(i\omega_n) - 
U n^f_i \delta_{ij}}{i\omega_n \delta_{ij}}
\end{eqnarray}
the unnormalized weight. $Z= \sum_{\{f\}} w_f^0$ is the 
``partition sum''. The determinant is to be taken over the spatial indices.
Given the weights, the new d-electron cluster Green function is given by
\begin{equation}
G_{ij}(z) = \sum_{\{f\}} w_f \left[{\cal{G}}^{-1}_{ij} (z) - U n_i^f
\delta_{ij} \right]^{-1}\, .
\end{equation}
The self--consistency loop closes by use of the Eqs. \ref{gbar},
\ref{gsck} and \ref{sigc}. 
\begin{figure}[h]
\leavevmode\centering\psfig{file=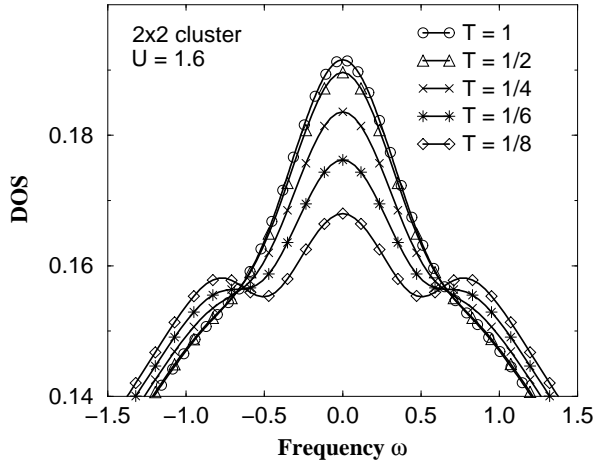,width=3.3 in}
\caption{Conduction electron DOS in the homogeneous phase
for various temperatures ($2\times 2$ cluster) and  $U=1.6$. Note
the emergence 
of ``charge transfer'' peaks with simultaneous suppression of the 2D van 
Hove peak at the band center. 
In contrast, the DMFA result is temperature independent.}
\label{dosfig1}
\end{figure}
Because the number of f-configurations grows exponentially with the 
cluster size we confine ourselves to $1\times1$, $2\times2$ and 
$4\times4$ clusters in 2D.  We first simultaneously determine the  
weights and the  Matsubara Green 
function. Then we use knowledge of the weights to find the retarded Green 
function. Convergence of the algorithm is fast for Matsubara frequencies, but 
relatively slow for real frequencies. 
Upon convergence we test the sum rules of the spectral function at 
the cluster momenta. 

The spectral 
functions are  always positive, and the sum rules for both the
cluster Green function as well as the host Green function
$\cal{G}$ are fulfilled within numerical accuracy 
for moderate interaction strength $U$. For large  $U$ a gap opens
in the DOS and convergence becomes more difficult for $\omega= 0$.
This is because the self energy for the momenta on the Fermi surface
(e.g. $\K =(\pi,0)$) approaches the atomic limit, 
$\Sigma(\omega) \approx U^2/4(\omega +i\eta)$ for frequencies inside the gap
($\eta$ is a positive infinitesimal).
This implies that as  $\omega \rightarrow 0$,  
$\mbox{Im}\Sigma\rightarrow -\infty$, which is rather difficult to converge
to. On the other hand, for all other frequencies the algorithm converges
to within the desired accuracy. Since the contribution to the DOS from 
$\omega=0$ is infinitesimal, the spectral sum rules are also fulfilled
to within numerical accuracy.
We emphasize that these peculiarities at $\omega= 0$ 
are only observed for the real frequency
algorithm.  For the  Matsubara frequency algorithm, 
the sum rules (which may be
re-expressed in terms of imaginary-time propagators) were always 
satisfied and the algorithm was perfectly stable 
(at least at the temperatures considered). 
\begin{figure}[h]
\leavevmode\centering\psfig{file=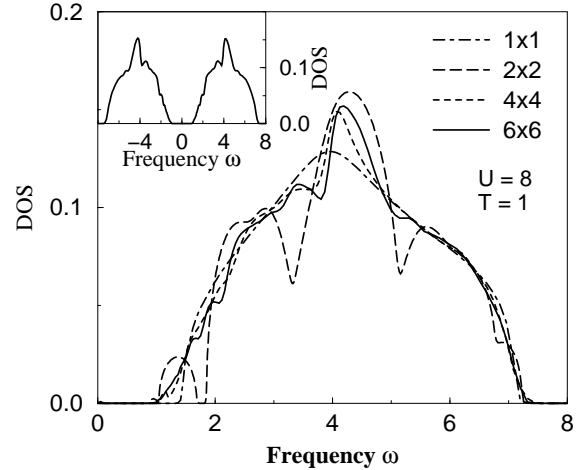,width=3.3 in}
\caption{Conduction electron DOS
in the homogeneous phase for various cluster sizes for  a fixed $T=1$ 
and $U=8$. Only half of the symmetric DOS is shown. Note that 
the artificial side band of the $2\times2$ cluster disappears
at larger cluster size.
The entire DOS for the $6\times 6$ cluster is shown in the inset.}
\label{dosfig2}
\end{figure}

In Fig. \ref{dosfig1} we show the DOS of the 
conduction electrons for the half filled case for
the $2\times 2$ cluster for $U=1.6$.
In DMFA there is {\it no} temperature dependence of the DOS, since 
 the weights of the unoccupied and occupied f--state are 
$w_o=w_1=1/2$, independent of temperature (in the homogeneous phase).
This is changed in the DCA, where the checkerboard configurations
begin to dominate as the temperature is lowered. The result is
the appearance of the "charge transfer" features in the DOS, the two peaks
separated by the interaction strength $U$.

Next, we explore the finite size effects of the DCA at large $U$ where the
DOS shows more features (including a gap), and finite size effects
are more severe.
In small clusters the  effects of periodic boundary conditions 
are strong.
Our results for the DOS at $U=8$ are shown in Fig. \ref{dosfig2}.
Notice how the spurious features of the $2\times 2$ cluster
(strong dips and an additional small gap) 
have essentially disappeared in the DOS of the  $4\times 4$ cluster, 
though small features at the edges of the gap remain (not discernible 
in the figure).
As larger clusters can not be evaluated 
exactly (too many configurations) we employ Monte Carlo sampling 
of the configurations. As a preliminary result of  work in progress 
we show the DOS of the $6\times 6$ cluster. 
Already at this modest cluster size all finite size features are essentially 
eliminated. This hints to the superior finite size scaling 
properties of the DCA as compared to the standard lattice techniques without
coupling to a host.

\begin{figure}[h]
\leavevmode\centering\psfig{file=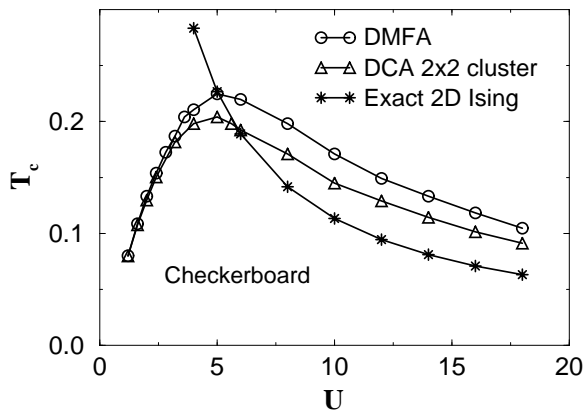,width=3.3 in}
\caption{Phase diagram of the 2D FKM at half filling. Compared to the
DMFA result (circles) $T_c$ of the 2x2 cluster DCA (triangles) 
is significantly suppressed for large interaction. At asymptotically large $U$
the  $T_c$ of the DCA
is bounded from below by the $T_c$ of the 2D Ising model.}
\label{tcvsu}
\end{figure}

Finally, we discuss the effect of nonlocal corrections on 
the transition temperature $T_c$ to the checkerboard phase.
Within the DCA, we find $T_c$ by estimating the temperature
where the order parameter in the broken symmetry phase  vanishes.
The phase diagram is displayed in Fig. \ref{tcvsu}.  
The nonlocal correlations of the DCA  suppress the $T_c$
compared to the DMFA estimate, except for 
weak $U$ where the nonlocal corrections to the vertex are very 
small (of order $U^2$ smaller than local contributions). For large $U$, 
however, the model maps onto an effective Ising model with a near-neighbor 
exchange coupling $J=t^2/2U$ and a corresponding
$T_c^{Ising} =1.134/U$\cite{peter2}.  
Fig. \ref{tcvsu} shows that already for the $2\times 2$ cluster
the achieved correction takes one almost half way
to the asymptotically ($U\rightarrow \infty$) exact $T_c$ of the 
2D Ising model.\\

{\it Conclusions} We have introduced a new dynamical cluster 
approximation that  includes
short--ranged spatial correlations in addition to the local 
correlations of the dynamical mean field approximation of strongly
interacting electron systems. The method interpolates between the
infinite lattice and the DMFA by evaluating the self energy on a 
finite size cluster with periodic boundary conditions. 
The DCA is a general scheme and is easily adapted to specific models
and various existing exact and perturbative techniques to solve 
these models. As an example we
applied the method to the Falicov--Kimball model in 2D and obtained
the DOS as a function of temperature for small cluster sizes. In addition,
we computed the critical temperature of the checkerboard phase transition
and showed that it is suppressed for large interactions when compared to the
result of DMFA.

Acknowledgments: It is a pleasure to acknowledge discussions with
J. Freericks, G.\ Baker, P.G.J.\ van Dongen, J. Gubernatis, A. Schiller 
and F.-C. Zhang.  This work was supported by NSF grants DMR-9704021 and 
DMR-9357199, and the Ohio Board of Regents Research Challenge Award 
(H.R.K.).  Travel support was  provided by  NATO (T.P., M.J.). Computer 
support was provided by the Ohio Supercomputer Center.



\begin{references}
\bibitem{note1} By causal we mean that the spectral function is positive 
semidefinite and that it preserves the sum rule.
\bibitem{comment} This is a simple consequence of the fact that for any
positive semi-definite function the Hilbert transform can be represented
as continued fraction with positive coefficients in the numerator of each of
the fractions.
\end{references}
\end{document}